\documentclass[epj]{svjour}
\usepackage{graphics}
\usepackage[utf8]{inputenc}
\usepackage{amsmath}
\usepackage{amssymb}
\usepackage[toc,page]{appendix}
\usepackage{geometry}
\usepackage{graphicx}
\usepackage{hyphenat}
\usepackage{graphicx} 
\usepackage{float}    
\begin{document}
\title{Third-quantized symmetry breaking and the selection of fundamental constants}
\author{Adam Balcerzak\inst{1,2} \and Konrad Marosek\inst{3}
}                     
\offprints{}          
\institute{Institute of Physics, University of Szczecin,
Wielkopolska 15, 70-451 Szczecin,  Poland \and Copernicus Center for Interdisciplinary Studies, Szczepa\'nska 1/5, 31-011 Krak\'ow, Poland \and Institute of Mathematics, Physics and Chemistry, Maritime University of Szczecin, Wa{\l }y Chrobrego 1-2, 70-500 Szczecin, Poland}
\date{Received: date / Revised version: date}
%
\abstract{
A mechanism is proposed in which the third-quantized wave function of the universe undergoes spontaneous symmetry breaking. In principle, this mechanism may be enabled by a potential arising from interactions between scalar and hypothetical fermionic degrees of freedom of the wave function in minisuperspace. A connection is established between the vacuum expectation value of the broken phase and the values of the speed of light and the gravitational constant. Crucially, a new, broken vacuum is established, which is linked to the symmetric state via Bogoliubov transformations. The resulting transformation coefficients indicate that this broken phase contains excitations relative to the symmetric vacuum. These excited modes are interpreted as the production of universes during a phase transition within the multiverse.
\PACS{
      {04.50.Kd}{Modified theories of gravity}   \and
      {04.60.-m}{Quantum gravity}
     } 
}
\titlerunning{Third-quantized symmetry breaking and fundamental constants}
\maketitle
\section{Introduction}
\label{intro}
Various attempts have been made to formulate a conceptually consistent theory with a varying speed of light (VSL). Apart from frameworks where the speed of light depends on the energy scale \cite{Amelino,Ellis,Ellis2,Amelino2,Kowalski}, most of these models entail a violation of Lorentz invariance, for which there is currently insufficient experimental evidence. Such theories typically assume a preferred frame, usually identified with the cosmological frame, wherein the model takes the form of a minimally coupled theory and the speed of light is linked to a scalar field \cite{Albrecht,Barrow1,Clayton,Clayton2}. Alternatively, bimetric VSL theories also violate Lorentz invariance by introducing two distinct metrics, one describing the geometry of spacetime and the other coupling to matter \cite{Drummond}.

In a specific VSL theory proposed in \cite{Magueijo1}, the speed of light is linked to a dynamical, non\hyp{}minimally coupled scalar field, preserving both the covariance and the local Lorentz invariance. This framework inspired a VSL model that additionally incorporates a varying gravitational constant, where both parameters are represented by non-minimally coupled scalar fields \cite{Balcerzak1}. In the canonically quantized case of this model, cosmogenesis occurs as a transition from a pre-big-bang to a post-big-bang phase of the evolution. The third-quantized counterpart of such a theory has also been considered \cite{Balcerzak2,Balcerzak3,Balcerzak4}, yielding the exact form of the third-quantized action. Interestingly, in the high-curvature limit, the form of the action takes a canonical form characterized by a kinetic and a constant mass term \cite{Balcerzak2}.

Quantum cosmology is based on the Wheeler-DeWitt equation, where the wave function is a complex scalar field. Many attempts have been made to introduce fermionic degrees of freedom into quantum cosmology. The most established approach relies on supersymmetry \cite{moniz1,moniz2,moniz3,moniz4,moniz5}. Another method relies on the procedure of taking the square root of the Wheeler-DeWitt equation \cite{Kim,Death,Yamazaki,Hojman,factor_ordering1}. This latter approach was motivated by several issues related to the Wheeler-DeWitt equation: the difficulties in providing a probabilistic interpretation, since the equation is of second order, and the ambiguities in factor ordering. An alternative method for introducing fermionic degrees of freedom into the wave function of the universe relies on the procedure of the Eisenhart-Duval lift \cite{Kan1,Kan2}. This approach relies on introducing an auxiliary variable to the original set of variables that constitutes the initial minisuperspace, which allows for the geometrization of the dynamics of the system, which are generally non-geometric in the presence of a potential. The dynamics on the extended minisuperspace are governed by an extended metric uniquely obtained in the procedure of the Eisenhart-Duval lift, providing a natural way to introduce a Dirac-like equation and thus fermionic degrees of freedom into the wave function of the universe. This approach was further explored in \cite{Balcerzak2023,Balcerzak2023_Cosmogenesis,Kucharski2025}, yielding novel scenarios of cosmogenesis and providing a framework for the elimination of type IV cosmological singularities. 

In \cite{Coleman1973}, it was demonstrated that radiative corrections can induce spontaneous symmetry breaking in theories whose tree-level approximations do not exhibit this behavior, which was shown for massless scalar electrodynamics and non-Abelian gauge theories. Moreover, when the effective potential is improved by the renormalization group equations, negative contributions from fermions to the renormalization parameter can cause the loop corrections to dominate the effective potential, which may subsequently result in the radiative breaking provided the potential has a stable minimum \cite{Masina2025}. Furthermore, the mechanism of radiatively induced breaking of the Grand Unified Theory (GUT) symmetry can lead to cosmological models in which certain cosmological puzzles are resolved \cite{Albrecht1982}.

The third-quantization procedure used in this work yields a formalism that describes the multiverse and its underlying processes. Under this framework, the Wheeler-DeWitt scalar wave function of the universe is promoted to an operator, analogous to the promotion of the Klein-Gordon field during field quantization \cite{Strominger1990,RoblesPerez2010,RoblesPerez2014}. The field operator subsequently acts to create and annihilate individual universes within the multiverse. Such formulation was adopted in \cite{Rubakov1988} to investigate the quantum creation of universes and the distribution of cosmological parameters. A similar third-quantized interaction scheme was also explored in \cite{McGuigan1990} to model decay processes between universes characterized by different cosmological constants. Another interesting consequence of introducing third quantization is the emergence of non-interacting negative-energy renormalization fields, which can resolve the problem of vacuum instability at the Planck scale, while providing a measurable modification to the Casimir effect and allowing for the possibility of an eternal universe \cite{Heintzmann}.

 In this paper, we consider a third-quantized action featuring a potential that admits spontaneous symmetry breaking (SSB) and investigate the physical consequences of this symmetry-breaking process. We argue that SSB in the third-quantized framework can be interpreted as a mechanism for the creation of a multiverse, in which the emergence of a nonvanishing vacuum expectation value of the third-quantized wave function plays a central role. Furthermore, we explore how this mechanism may provide a dynamical origin for the values of fundamental constants, in particular the speed of light and the gravitational constant, through their dependence on the vacuum expectation value generated by spontaneous symmetry breaking.

Our paper is organized as follows. In Sec.~\ref{sec:1}, we introduce the third-quantized framework and describe the mechanism of spontaneous symmetry breaking, showing how the resulting nonvanishing vacuum expectation value can be associated with the creation of a multiverse. In Sec.~\ref{sec:2}, we investigate how the vacuum selected through spontaneous symmetry breaking determines the values of the fundamental constants, in particular the speed of light and the gravitational constant. In Sec.~\ref{conc}, we give our conclusions.

\section{Cosmogenesis via spontaneous symmetry breaking}
\label{sec:1}
The third quantized action for the varying $c$ and $G$ model defined with the classical action (\ref{actionBD}) (see App. (\ref{app:1})) is given by \cite{Balcerzak1,Balcerzak2,Balcerzak3,Balcerzak4}:
\begin{equation}
\label{3action}
S[\Psi]=\frac{1}{2}\int dT d^{2}X\left[  (\partial_T{\Psi})^2 - (\nabla \Psi)^2 -m_{eff}^2\Psi^2 \right],
\end{equation}
where $\nabla$ is a two dimensional gradient operator associated with the spatial dimensions of the minisuperspace while $\partial_T$ denotes the differentiation with respect to the time variable. The corresponding field equation is:
\begin{equation}
\left(\partial_T^2-\nabla^2+m_{eff}^2\right)\Psi=0.
\end{equation}
As already mentioned, a radiatively induced mechanism - driven by quantum loop corrections to the effective potential - can, in principle, destabilize the symmetric vacuum and lead to spontaneous symmetry breaking. At the phenomenological level, this transition is described by a third-quantized action:
\begin{equation}
S[\Psi] = \int\!dT d^2X \Big[\! \tfrac{1}{2}(\partial_T\Psi)^2 \!-\! \tfrac{1}{2}(\nabla\Psi)^2 \!+\! \tfrac{1}{2}\mu^2\Psi^2 \!-\! \tfrac{\lambda}{4}\Psi^4 \!\Big]
\end{equation}
where $\mu^2>0$ and $\lambda>0$.  The corresponding potential is:
\begin{equation}
\label{V_ssb}
V(\Psi)=-\frac12\mu^2\Psi^2+\frac{\lambda}{4}\Psi^4.
\end{equation}

Before the symmetry breaking, the dynamics is given by the free third-quantized action,
\begin{equation}
S[\Psi]=\frac12\int dT d^2 X\left[(\partial_T\Psi)^2-(\nabla\Psi)^2-m^2\Psi^2\right],
\end{equation}
which yields the following Klein-Gordon equation $(\partial_T^2-\nabla^2+m^2)\Psi=0$.
The field can be expanded into normal modes,
$\Psi=\sum_\mathbf{k} \left[a_\mathbf{k} u_\mathbf{k}+a_\mathbf{k}^\dagger u_\mathbf{k}^*\right]$, where the mode functions $u_k$ fulfill the Klein-Gordon equation. Separating variables through $u_k=f_k(T)e^{i\mathbf{k}\cdot\mathbf{X}}$ gives the harmonic equation $\ddot f_k+\omega_k^2 f_k =0$ with dispersion relation $\omega_k =\sqrt{k^2+m^2}$. Selecting positive-frequency solutions yields
\begin{equation}
u_k=\frac{1}{\sqrt{2\omega_k}} e^{-i\omega_kT+i\mathbf{k}\cdot\mathbf{X}}.
\end{equation}
The creation and annihilation operators $a_k$ and  $a_k^\dagger$ satisfy the standard commutation relations with the vacuum $\vert{}0\rangle$ defined by $a_k\vert{}0\rangle=0$.

Upon the symmetry breaking, the field acquires a non-zero expectation value $\langle\Psi\rangle = \pm\sqrt{\mu^2/\lambda}\equiv v$, determined by the condition $dV/d\Psi=0$. Expanding the field around one of these minima gives
\begin{equation}
\Psi=v+h.
\end{equation}
The Hamiltonian density expanded around the vacuum $v = \sqrt{\mu^2/\lambda}$ after the symmetry breaking is
\begin{align}
\label{eq:H_density_split}
\mathcal{H}_{\text{density}} &= \mathcal{H}_0 + \mathcal{H}_1 + \mathcal{H}_2 + \mathcal{H}_3 + \mathcal{H}_4,
\end{align}
where 
\begin{align}
\mathcal{H}_0 &= -\frac{\mu^4}{4\lambda}, \qquad \mathcal{H}_1 = 0, \label{eq:H0_H1} \\
\mathcal{H}_2 &= \frac{1}{2}(\partial_T h)^2 + \frac{1}{2}(\nabla h)^2 + \frac{1}{2} m_h^2 h^2, \label{eq:H2} \\
\mathcal{H}_3 &= \frac{\sqrt{\lambda} m_h}{\sqrt{2}} h^3, \qquad \mathcal{H}_4 = \frac{\lambda}{4} h^4, \label{eq:H3_H4}
\end{align}
with $m_h = \sqrt{2}\mu$. In the limit of small perturbations ($h \ll v$), the field $h(T,X)$ satisfies the Klein-Gordon equation
\begin{equation}
\left( \partial_T^2 - \nabla^2 + m_h^2 \right) h(T,X) = 0, \label{eq:KG_fluctuation}
\end{equation}
with the new mass $m_h = \sqrt{2}\mu$.

The field $h$ can now be expanded into normal modes as $h=\sum_\mathbf{k} \left[b_\mathbf{k} v_\mathbf{k}+b_\mathbf{k}^\dagger v_\mathbf{k}^*\right]$, where the mode functions $v_k$ satisfy the Klein-Gordon equation with the shifted mass $m_h = \sqrt{2}\mu$. Separating variables via $v_k(T,\mathbf{X})=g_k(T)e^{i\mathbf{k}\cdot\mathbf{X}}$ yields $\ddot g_k+\Omega_k^2 g_k=0$, leading to the modified dispersion relation $\Omega_k=\sqrt{k^2+m_h^2}$. Selecting positive-frequency solutions gives
\begin{equation}
v_k(T,\mathbf{X})=\frac{1}{\sqrt{2\Omega_k}}e^{-i\Omega_kT+i\mathbf{k}\cdot\mathbf{X}}.
\end{equation}
The new annihilation and creation operators, $b_k$ and $b_k^\dagger$, define the symmetry-broken vacuum state $|0_h\rangle$ via $b_k|0_h\rangle=0$.

We consider a sudden symmetry breaking occurring at $T=0$, after which the field acquires the vacuum expectation value $\langle \Psi \rangle = v$. As a result, the physical fluctuation $h(T,\mathbf{X})$ gains a mass $m_h = \sqrt{2}\mu$. We further assume continuity conditions of the form:
\begin{align}
&\Psi(0^-, \mathbf{X}) - v = h(0^+, \mathbf{X}), \label{eq:field_cont} \\
&\partial_T \Psi(0^-, \mathbf{X}) = \partial_T h(0^+, \mathbf{X}). \label{eq:deriv_cont}
\end{align}
The corresponding Bogoliubov transformations take the form
\begin{align}
b_{\mathbf{k}} &= \alpha_k a_{\mathbf{k}} + \beta_k^* a_{-\mathbf{k}}^\dagger, \label{eq:b_op} \\
b_{\mathbf{k}}^\dagger &= \alpha_k^* a_{\mathbf{k}}^\dagger + \beta_k a_{-\mathbf{k}}, \label{eq:bdag_op}
\end{align}
where $\alpha_k$ and $\beta_k$ are given by
\begin{align}
\alpha_k &= \frac{1}{2} \left( \sqrt{\frac{\Omega_k}{\omega_k}} + \sqrt{\frac{\omega_k}{\Omega_k}} \right), \label{eq:alpha_final} \\[6pt]
\beta_k &= \frac{1}{2} \left( \sqrt{\frac{\Omega_k}{\omega_k}} - \sqrt{\frac{\omega_k}{\Omega_k}} \right). \label{eq:beta_final}
\end{align}
These coefficients explicitly satisfy the canonical unitarity relation $|\alpha_k|^2 - |\beta_k|^2 = 1$.

In terms of the original vacuum state, the expectation value of the ``broken-phase'' number operator $N_\mathbf{k}=b_\mathbf{k}^\dagger b_\mathbf{k}$ evaluates to
\begin{equation}
N_\mathbf{k}=|\beta_k|^2=\frac14\left(\sqrt{\frac{\Omega_k}{\omega_k}}-\sqrt{\frac{\omega_k}{\Omega_k}}\right)^2.
\end{equation}
The total number density of universes is
\begin{equation}
n= \int_0^\infty \frac{k\,dk}{2\pi} N_k.
\end{equation}
\begin{figure}[ht]
\centering
\includegraphics[width=0.48\textwidth]{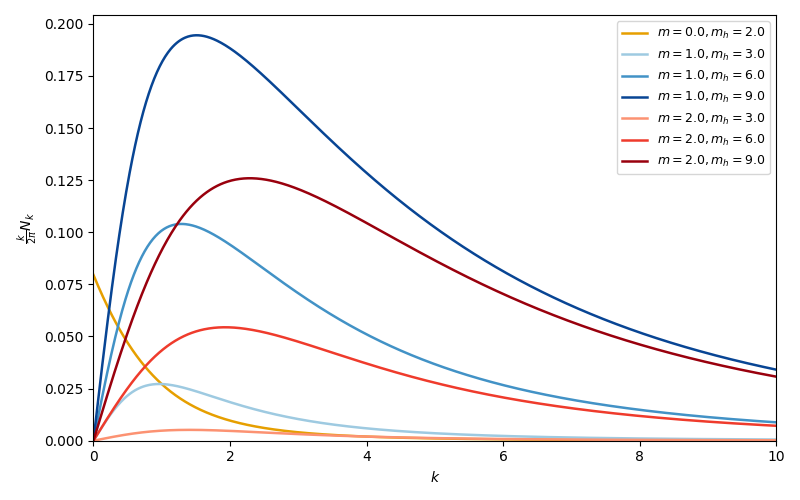}
\caption{Integrand of the total universe number density.}
\label{fig:density}
\end{figure}

The resulting differential particle creation spectrum, $\frac{k}{2\pi} N_k$, exhibits distinct physical regimes depending on the initial field mass $m$ and the post-transition mass $m_h$ (Fig.~\ref{fig:density}). In the infrared limit ($k \to 0$), the behavior depends critically on whether the initial state is massive or massless. For $m = 0$, the mode occupation diverges as $N_k \propto k^{-1}$, which exactly balances the linear phase-space factor $k$ to yield a non-zero constant origin, $\lim_{k \to 0} \frac{k}{2\pi} N_k = \frac{m_h}{8\pi}$. Conversely, for any $m > 0$, the zero-mode occupation $N_0$ remains finite, causing the explicit factor of $k$ to suppress production and force all massive curves to vanish at $k = 0$. At intermediate momenta, the magnitude of the mass jump $\Delta m = m_h - m$ dictates the overall amplitude and shape of the spectrum. Larger mass discontinuities inject more energy into the system, increasing the peak spectral height. Furthermore, for a fixed final mass $m_h$, smaller initial masses $m$ result in significantly enhanced universes creation across all modes. Finally, in the ultraviolet limit ($k \gg m_h$), all spectra decay to zero as $k^{-3}$. High-momentum modes possess kinetic energies far exceeding the mass scale of the transition, making them effectively insensitive to the sudden change.

\section{Fundamental constants and the spontaneously broken vacuum}
\label{sec:2}

In the mechanism of spontaneous symmetry breaking, the third-quantized wave function of the universe acquires a non-zero vacuum expectation value, $\langle \Psi \rangle = \sqrt{\mu^2 / \lambda}$. Consequently, the mass term shifts to:
\begin{equation}
m_{\text{eff}} = \sqrt{2\lambda} \langle \Psi \rangle
\end{equation}
Using Eqs. (\ref{eq:c})–(\ref{eq:al}), we relate the initial values of the fundamental constants to the vacuum expectation value of the third-quantized wave function:
\begin{equation}
c_{\text{init}} = C_{\text{final}} \cdot \left(\sqrt{2\lambda}\langle \Psi \rangle\right)^{-\alpha_c},
\end{equation}
\begin{equation}
G_{\text{init}} = G_{\text{final}} \cdot \left(\sqrt{2\lambda}\langle \Psi \rangle\right)^{\gamma_g},
\end{equation}
\begin{equation}
\alpha_{\text{EM, init}} = \alpha \cdot \left(\sqrt{2\lambda}\langle \Psi \rangle\right)^{\alpha_c},
\end{equation}
\begin{equation}
H_{\text{init}} = h \cdot \left(\sqrt{2\lambda}\langle \Psi \rangle\right)^{-\alpha_h},
\end{equation}
\begin{equation}
l_{\text{P, init}}  = l \left( \sqrt{2\lambda}\langle \Psi \rangle\right)^\sigma,
\end{equation}
where explicit expressions for the constants $\alpha_c$, $\gamma_g$, $\alpha_h$, and $\sigma$ are provided in the Appendices.
Spontaneous symmetry breaking establishes the vacuum expectation values governing fundamental parameters while simultaneously specifying the initial conditions for cosmic inflation. Consequently, the determination of physical constants and the initiation of inflation emerge from a single phase transition rather than distinct cosmological mechanisms.

\section{Conclusions}
\label{conc}
We proposed a mechanism based on spontaneous symmetry breaking in which the broken vacuum gives rise to a population of universes. Furthermore, we link the values of the fundamental constants to the expectation value of the third-quantized wave function in the symmetry-broken vacuum state. Through non-trivial Bogoliubov transformations between the unbroken and broken vacuum states, the field transition induces a non-zero occupation number of universe states, giving rise to the third-quantized condensate.

On the classical level, where cosmological evolution stems purely from general relativity, the high-curvature regime yields an intriguing limit as the Planck length vanishes. In this limit, gravitational interactions decouple and spacetime remains smooth even as the curvature approaches infinity. Spontaneous symmetry breaking prevents this behavior by dynamically fixing the fundamental constants - including the Planck length - at the moment of the phase transition, which marks the appearance of the condensate in the multiverse.

A natural byproduct of this mechanism is the generation of initial conditions for cosmological inflation. As the third-quantized wave function acquires mass during spontaneous symmetry breaking, the Hubble parameter along with the remaining fundamental constants are set at the symmetry-breaking scale, which defines the expansion rate during inflation.

Third quantization shows that spontaneous symmetry breaking replaces the classical singular limit with a physical phase transition. By setting both the fundamental constants and the Hubble parameter during this transition, the symmetry-breaking mechanism provides a clear field-theoretic source for the initial inflation rate. This demonstrates that key cosmological parameters can in priciple be calculated directly from the quantum properties of the broken vacuum rather than assumed as free inputs.

\section*{Acknowledgements}
We gratefully acknowledge the use of Gemini in refining the manuscript’s language, structure, and overall clarity, as well as in improving the presentation and formatting of the equations.

\appendix

\renewcommand{\theequation}{A.\arabic{equation}}
\renewcommand{\thefigure}{A\arabic{figure}}

\setcounter{equation}{0}
\setcounter{figure}{0}

\begin{appendices}

\section{Some results from the non-minimally coupled varying $c$ and $G$ cosmological model}
\label{app:1}
The covariant and locally Lorentz-invariant form of the action for the non-minimally coupled theory with varying $c$ and $G$ is given by \cite{Balcerzak1,Balcerzak2,Balcerzak3,Balcerzak4}:
\begin{equation}
\label{actionBD}
S \!=\! \int \!\sqrt{-g}\left[ \delta (R+\Lambda) \!+\! \frac{\omega}{2}\frac{\partial_\mu \delta \partial^\mu \delta}{\delta} \!+\! \delta \partial_\mu \beta \partial^\mu \beta\right] \! d^4x
\end{equation}
where $\beta$ and $\delta$ are some scalar fields. The relation between $c$ and $G$ and the scalar fields $\beta$ and $\delta$ are
\begin{equation} 
\label{eq:cG}
c^3 = e^\phi, \quad G = e^\psi,
\end{equation}
where 
\begin{equation} 
\label{eq:phi_psi_def}
\phi = \frac{\beta}{\sqrt{2\omega}} + \frac{1}{2} \ln \delta, \quad \psi = \frac{\beta}{\sqrt{2\omega}} - \frac{1}{2}\ln\delta.
\end{equation}
This specific action is inspired by the covariant and locally Lorentz-invariant varying speed of light theories introduced in \cite{Magueijo1}.

The explicit solutions for the scale factor $a(x^0)$ and the field $\delta(x^0)$ are given by:
\begin{equation}
\label{eq:a}
a(x^0) = D^{-2} \cdot e^{-2F \cdot x^0} \cdot \sinh^{-M} | l \cdot x^0|
\end{equation}
\begin{equation}
\label{eq:delta}
\delta(x^0) = D^6 \cdot e^{6F \cdot x^0} \cdot \sinh^{-W} | l \cdot x^0|
\end{equation}
where $l \equiv \sqrt{(A^2-9)\Lambda}$ while $M=\frac{3-A^2}{9-A^2}$ and $W= \frac{2A^2}{9-A^2}$.
The expression for $\beta$ is:
\begin{equation}
\label{eq:beta_sol}
\beta = B_1 \cdot x^0 + B_0,
\end{equation}
where $B_1$ and $B_0$ are the integration constants.

The proper time experienced by a comoving observer is defined by:
\begin{equation}
\label{eq:dtau_def}
d\tau = \frac{d\bar{x}^0}{c(x^0)}
\end{equation}
where the rescaled cosmic time element $d\bar{x}^0$ is constrained by the gauge choice $N = a^3\delta$:
\begin{equation}
\label{eq:dxbar_def}
d\bar{x}^0 = a^3 \delta \, dx^0.
\end{equation}

\section{Derivation of the speed of light}
\label{app:2}
By combining (\ref{eq:beta_sol}) the logarithm of $\delta(x^0)$, the field $\phi(x^0)$ can be expressed as:
\begin{equation}
\phi(x^0) = \phi_0 + K_\phi \cdot x^0 - \frac{W}{2} \ln | \sinh(l \cdot x^0) |.
\end{equation}
where $\phi_0$ and $K_\phi$ are consolidated constants. From the first formula in  (\ref{eq:cG}) we get:
\begin{equation}
c(x^0) = \exp\left( \frac{\phi_0}{3} + \frac{K_\phi}{3}x^0 - \frac{W}{6}\ln|\sinh(l \cdot x^0)| \right).
\end{equation}
Defining the new consolidated parameters $c_0 \equiv e^{\phi_0/3}$ and $K \equiv K_\phi / 3$, we find:
\begin{equation}
\label{eq:c_final}
c(x^0) = c_0 \cdot e^{K \cdot x^0} \cdot \sinh^{-\frac{W}{6}} | l \cdot x^0|.
\end{equation}
From Eqs.~(\ref{eq:c_final}), (\ref{eq:dxbar_def}), (\ref{eq:a}), (\ref{eq:delta}), and~(\ref{eq:dtau_def}), the differential of the physical proper time $\tau$ is given by:
\begin{equation}
\label{eq:dtau}
d\tau = \frac{1}{c_0} \cdot e^{-K \cdot x^0} \cdot \sinh^{\frac{W}{6} - 1} | l \cdot x^0| \, dx^0.
\end{equation}
Equation~(\ref{eq:dtau}) can be integrated in the limit $x^0 \to \infty$, which corresponds to the Big Bang singularity, yielding:
\begin{equation}
\label{eq:tau_lim}
\tau = \frac{2^{1 - \frac{W}{6}}}{c_0 \cdot (-\Omega)} \cdot e^{\Omega \cdot x^0_i},
\end{equation}
where
\begin{equation}
\label{eq:omega_def}
\Omega \equiv l \left( \frac{W}{6} - 1 \right) - K.
\end{equation}
Since $\Omega < 0$ the physical proper time $\tau$ experienced by a comoving observer is strictly finite.

The effective mass parameter of the cosmological model is \cite{Balcerzak1}:
\begin{equation}
m_{\text{eff}}^2 = \Lambda e^{-\frac{2}{r}\eta},
\end{equation}
where $r=\sqrt{\frac{V_0}{A^2-9}}$.
From the explicit Hamiltonian solutions $\eta(x^0) = \ln \sinh | l x^0|$ the effective mass formula yields \cite{Balcerzak2}:
\begin{equation}
\label{eq:meff_exact}
m_{\text{eff}}^2(x^0) = \frac{\Lambda}{\sinh^{\frac{2}{r}} | l \cdot x^0|}
\end{equation}
Substituting  $x^0(\tau) = \frac{1}{|\Omega|} \ln\left(\frac{C}{\tau}\right)$ (the inverse of \ref{eq:tau_lim}), where $C$ is a positive constant, into (\ref{eq:meff_exact}) taken in the limit ($x^0 \to \infty$) one gets:
\begin{equation}
\label{eq:meff_final}
m_{\text{eff}}^2(\tau) \approx M_0 \tau^{\frac{2l}{r \cdot |\Omega|}},
\end{equation}
where $M_0 \equiv \Lambda \cdot 2^{2/r}  C^{-\frac{2l}{r|\Omega|}}$. Inverting (\ref{eq:meff_final}) one gets:
\begin{equation}
\label{eq:tau_of_meff}
\tau(m_{\text{eff}}) = T_0 \cdot m_{\text{eff}}^{\frac{r \cdot |\Omega|}{l}},
\end{equation}
where $T_0 \equiv (1/M_0)^{\frac{r|\Omega|}{2l}}$.
Since $r$, $l$, and $|\Omega|$ are strictly positive numbers, the net exponent is positive. 
This proves that as the effective scattering barrier vanishes ($m_{\text{eff}} \to 0$), the physical proper time tends to zero ($\tau \to 0$).

The speed of light $c$ can now be derived directly as a function of the effective potential mass parameter $m_{\text{eff}}$ in the near-singularity regime ($x^0 \to \infty$).
By using (\ref{eq:c_final}), the inverse of (\ref{eq:tau_lim}) and (\ref{eq:tau_of_meff}) we finally get
\begin{equation}
\label{eq:c_fin}
c(m_{\text{eff}}) \approx C_{\text{final}} \cdot m_{\text{eff}}^{-\frac{r \left(K - \frac{Wl}{6}\right)}{k}}, 
\end{equation}
where $C_{\text{final}} = c_0 \cdot 2^{\frac{W}{6}} \cdot \left(\frac{C}{T_0}\right)^{\frac{6K - Wl}{6|\Omega|}}$.

Under the model constraints ($A^2 > 9$), the scaling term $W = \frac{2A^2}{9-A^2}$ is strictly negative. Consequently, the term $-\frac{Wl}{6}$ acts as a positive addition, ensuring the bracketed coefficient $\left(K - \frac{Wl}{6}\right)$ remains positive. With the leading minus sign, the overall exponent is strictly negative. This proves that as the effective mass potential barrier collapses to zero ($m_{\text{eff}} \to 0$), the physical speed of light scales as an inverse power law and diverges directly to infinity ($c \to \infty$) at the moment of the Big Bang singularity.

\section{Asymptotic derivation of the gravitational constant}
Similarily, the gravitational constant expressed in terms of the effective mass, $G(m_{\text{eff}})$, is:
\begin{equation}
\label{eq:G_fin}
G(m_{\text{eff}}) \approx G_{\text{final}} \cdot m_{\text{eff}}^{-\frac{r \cdot \left(3K - 6F + \frac{Wk}{2}\right)}{k}},
\end{equation}
where
\begin{equation}
G_{\text{final}} \equiv G_0 \cdot (C \cdot T_0)^{-\frac{3K - 6F + \frac{Wk}{2}}{|\Omega|}}.
\end{equation}
The the expression $\left(3K - 6F + \frac{Wk}{2}\right)$ evaluates to a net negative value. The overall net exponent becomes strictly positive, which proves that as the effective potential mass barrier goes to zero ($m_{\text{eff}} \to 0$), the physical gravitational constant vanishes ($G \to 0$) at the Big Bang singularity.

\section{Definition of the Non-Degenerate parameters}
The asymptotic expressions for the speed of light (\ref{eq:c_fin}) and the gravitational constant (\ref{eq:G_fin}) contain degenerate combinations of background parameters and integration constants. To resolve this degeneracy, the system is mapped onto exactly four mutually independent parameters: $C_{\text{final}}$, $G_{\text{final}}$, $\alpha_c \equiv \frac{r \left(K - \frac{Wk}{6}\right)}{k} > 0$ and $\gamma_g \equiv \frac{r \left(6F - 3K - \frac{Wk}{2}\right)}{k} > 0$.
Consequently, the fundamental constants are expressed as power-law functions of the new independent parameters:
\begin{equation}
c(m_{\text{eff}}) = C_{\text{final}} \cdot m_{\text{eff}}^{-\alpha_c}
\end{equation}
\begin{equation}
G(m_{\text{eff}}) = G_{\text{final}} \cdot m_{\text{eff}}^{\gamma_g}
\end{equation}

\section{Extended Fundamental Parameters Scaling in the High-Curvature Limit}
In addition to the expressions for $c(m_{\text{eff}})$ and $G(m_{\text{eff}})$, the system defines the scaling functions for the Planck length $l_P(m_{\text{eff}})$, the Hubble parameter $H(m_{\text{eff}})$, and the fine-structure constant $\alpha_{\text{EM}}(m_{\text{eff}})$:
\begin{align}
\label{eq:c}
c(m_{\text{eff}}) &= C_{\text{final}} \cdot m_{\text{eff}}^{-\alpha_c} \\
\label{eq:G}
G(m_{\text{eff}}) &= G_{\text{final}} \cdot m_{\text{eff}}^{\gamma_g} \\
\label{eq:l}
l_P(m_{\text{eff}}) &= l \cdot m_{\text{eff}}^{\sigma} \quad \left(\text{where } \sigma = \frac{\gamma_g + 3\alpha_c}{2}\right) \\
\label{eq:H}
H(m_{\text{eff}}) &= h \cdot m_{\text{eff}}^{-\alpha_h} \quad \left(\text{where } \alpha_h = \frac{r}{k} |\Omega|\right) \\
\label{eq:al}
\alpha_{\text{EM}}(m_{\text{eff}}) &= \alpha \cdot m_{\text{eff}}^{\alpha_c}
\end{align}
The system is specified by the independent parameters $C_{\text{final}}$, $G_{\text{final}}$, $h$, $\alpha_c$, $\gamma_g$, and $\alpha_h$.

\end{appendices}

\end{document}